\documentclass[a4paper, 11 pt]{article}
\usepackage[T1]{fontenc}
\usepackage[utf8]{inputenc}
\usepackage[english]{babel}
\usepackage{amsmath}
\usepackage{amssymb}
\usepackage{braket}
\usepackage{mathtools}
\usepackage{dsfont}
\usepackage{subcaption}
\usepackage{booktabs}
\usepackage{yfonts}
\usepackage{setspace}
\usepackage{cite}
\usepackage{verbatim}

\usepackage{amstext} 
\usepackage{array}   
\newcolumntype{C}{>{$}c<{$}} 

\usepackage{geometry}
\usepackage[usenames,dvipsnames]{color}
\usepackage{hyperref}
\hypersetup{
  colorlinks,
  citecolor=Blue,
  linkcolor=Blue,
  urlcolor=Blue}

\numberwithin{equation}{section}

\def\be{\begin{equation}}
\def\ee{\end{equation}}

\newcommand{\diff}{\mathrm{d}}

\def\qq{{\hat q}}
\def\vv{v}
\def\ww{\mathsf{w}}

\begin{document}

\pagestyle{empty}

\begin{center}

$\,$
\vskip 0.5cm

{\LARGE{\bf BMPV black hole at first order in $\boldsymbol{\alpha'}$}}

\vskip 1cm

Alejandro Ruip\'erez

\vskip 1cm

\end{center}

\renewcommand{\thefootnote}{\arabic{footnote}}

\begin{center}
{\it Dipartimento di Fisica e Astronomia ``Galileo Galilei'', Universit\`a di Padova,\\ Via Marzolo 8, 35135, Padova, Italy\\[2mm]
INFN Sezione di Padova, Via Marzolo 8, 35135, Padova, Italy}

\vskip 3cm

 {\bf Abstract} 
\end{center}

{\noindent We consider the low-energy effective action of the heterotic string and derive an analytic solution describing the first-order $\alpha'$ corrections to the supersymmetric and extremal BMPV black hole with three unequal charges. The solution interpolates between an asymptotically-flat region and the near-horizon geometry. We compute the corrected black hole entropy using a generalization of Wald formula available in the literature, which correctly accounts for the Lorentz Chern-Simons term. The resulting expression agrees with recent results in the literature, which are based on the evaluation of an appropriate supersymmetric index.}

\newpage
\setcounter{page}{1}
\pagestyle{plain}

\tableofcontents

\vskip 1cm

\section{Introduction}

One of the key challenges for any theory of quantum gravity is to provide a microscopic derivation of black hole entropy. In string theory, this was originally achieved by Strominger and Vafa, who showed that the Bekenstein-Hawking entropy of certain supersymmetric black holes in five dimensions is reproduced by the statistical entropy of the underlying microscopic system \cite{ Strominger:1996sh}. Soon after, this was extended to account for rotation \cite{Breckenridge:1996is}, and to related black holes in four dimensions \cite{Maldacena:1997de}. These entropy matches were originally achieved in a regime where quantum-gravitational corrections can be safely ignored. The incorporation of quantum effects, particularly in supersymmetric settings where enhanced control is possible, naturally emerged as one of the next major steps and became a central goal of the field in the coming decades.

On the gravity side, we can distinguish between two types of corrections to the semiclassical Bekenstein-Hawking formula. The first arises from quantum fluctuations of massless fields in the black hole background. The second, which will be the focus of this paper, is due to the presence of higher-derivative corrections in the low-energy effective action.  In contrast to the former, the latter are sensitive to the ultraviolet completion of the theory. In the context of string theory, these arise after integrating out massive string states, with masses above the string scale $1/\sqrt{\alpha'}$. As it is well known, this gives rise to an \textit{a priori} infinite series of higher-derivative terms, to which we commonly refer as the $\alpha'$ expansion. In most situations of interest, the dimensionless expansion parameter in this series is the characteristic curvature scale of the background in string units. This means that these stringy effects become relevant only when the finite size of the black hole is taken into account, leading to corrections to both the geometry and the thermodynamic properties. Most notably, in presence of higher-curvature interactions, the entropy is no longer given by the Bekenstein--Hawking formula, but by the generalization due to Wald \cite{Wald:1993nt,Iyer:1994ys}.

The study of $\alpha'$ corrections presents several technical challenges. To begin with, our knowledge of $\alpha'$ corrections in string effective actions is by no means complete. Although in principle the underlying microscopic theory contains all the information required to determine the full tower of higher-derivative corrections, extracting them in practice is rather involved. As a consequence, only a few terms in the $\alpha'$ expansion of the various string theories are explicitly known. The case which is perhaps best understood is the heterotic string, whose complete effective action and supersymmetry transformations were presented in \cite{Bergshoeff:1989de} up to order $\alpha'^3$, extending earlier results \cite{Gross:1986mw, Metsaev:1987zx}. Even when the relevant corrections are known, solving the resulting corrected equations of motion and interpreting the corresponding solutions remains a formidable task. In particular, often one has to deal with subtleties in the definition of conserved charges, which have been source of confusion in the past. 

Several strategies to overcome these difficulties were proposed in the literature over the years. In particular, the application of superconformal calculus techniques \cite{Bergshoeff:1980sw, deWit:1980lyi, Freedman:2012zz} to construct higher-derivative supersymmetric invariants allowed to make progress in computing corrections to the entropy of asymptotically-flat BPS black holes in four and five dimensions. The four-dimensional case was investigated shortly after \cite{Strominger:1996sh, Breckenridge:1996is, Maldacena:1997de}, starting with \cite{Behrndt:1996jn,Behrndt:1998eq, LopesCardoso:1998tkj, LopesCardoso:1999fsj, LopesCardoso:2000qm, Mohaupt:2000mj}. This benefited from the fact that the supersymmetrization of the Weyl-squared term in $N=2$ supergravity was already known at that time \cite{Bergshoeff:1980is}. In contrast, progress in five dimensions had to wait a few years \cite{Castro:2007hc, Castro:2007ci, Castro:2008ne, Castro:2008ys, deWit:2009de}, being mostly triggered by the construction of a supersymmetric $R^2$ invariant in \cite{Hanaki:2006pj}.\footnote{Earlier results which are not based on the supersymmetric $R^2$ invariant constructed in \cite{Hanaki:2006pj} appeared in \cite{Guica:2005ig}.}
In the meantime, Sen developed the entropy function formalism, providing an efficient method for computing corrections to the entropy of extremal black holes ---supersymmetric or not--- which does not require explicit knowledge of the corrected solution \cite{Sen:2005wa, Sen:2005iz, Sahoo:2006pm, Sen:2007qy, Sen:2008vm}.\footnote{Here, as well as in the rest of the paper, we are assuming that we are treating the higher-derivative corrections perturbatively.} Eventually, this boils down to the fact that the two-derivative solution extremizes the (two-derivative part of the) action. Thus, the fact that the two-derivative solution suffices to extract the corrected thermodynamics must hold in general ---not just for extremal black holes---, as it was pointed out in \cite{Reall:2019sah}. In the last years, this observation has been exploited to compute higher-derivative corrections to the thermodynamics of supersymmetric AdS${}_5$ black holes \cite{Bobev:2022bjm,Cassani:2022lrk,Cassani:2024tvk}, allowing for a precise holographic match of the corresponding CFT results beyond leading order in the large-$N$ expansion \cite{Cassani:2021fyv,Cassani:2024tvk}.\footnote{See also \cite{Bobev:2020egg, Bobev:2021oku} for earlier results for AdS${}_4$ black holes.} \footnote{These results have been very recently re-derived using  equivariant localization techniques in supergravity in \cite{BenettiGenolini:2026qdm, Gaar:2026nqq}, without even making explicit use of the two-derivative solutions. Another closely related approach for obtaining these corrections in a remarkably efficient manner is through the equivariant integration of the anomaly polynomial. The power of this method, already employed in \cite{Ohmori:2021dzb,Cassani:2024tvk}, has been demonstrated most strikingly in \cite{Cassani:2026teb}, where it has been used to derive the on-shell action of supersymmetric black holes with non-trivial horizon topology in five dimensions, with both flat and AdS asymptotics, including higher-derivative corrections.}

In comparison with the development of techniques for computing  corrections to the entropy and charges, little progress was made in constructing the corresponding $\alpha'$-corrected solutions beyond the near-horizon limit. Restricting our attention to the supersymmetric case, the first-order $\alpha'$ corrections to the heterotic version of the ``Strominger--Vafa''  and ``Maldacena--Strominger--Witten'' black holes were obtained in \cite{Cano:2018qev,Chimento:2018kop,Cano:2018brq}, more than two decades after the corresponding two-derivative solutions had been constructed \cite{Cvetic:1995uj,Cvetic:1996xz} and their entropy successfully reproduced microscopically \cite{Strominger:1996sh,Breckenridge:1996is, Maldacena:1997de}.\footnote{The $\alpha'$ corrections to the finite-temperature black holes with three and four charges were later studied in \cite{Cano:2022tmn, Zatti:2023oiq}.} Subsequently, the two-charge case ---namely, the so-called small black holes--- was investigated in \cite{Cano:2018hut, Ruiperez:2020qda, Cano:2021dyy, Massai:2023cis}. Having analytic solutions at hand is always advantageous, but it proved particularly valuable in the latter case, where it enabled a reassessment of earlier claims in the literature regarding the resolution of the singular horizon of small black holes by $\alpha'$ corrections \cite{Sen:1995in, Dabholkar:2004yr}. Concretely, the conclusion of \cite{Cano:2018hut, Ruiperez:2020qda, Cano:2021dyy, Massai:2023cis} is that heterotic small black holes remain singular after including the first-order $\alpha'$ corrections. Moreover, it was argued that the black holes whose entropy was claimed to match the microscopic degeneracy of the two-charge system \cite{Dabholkar:1989jt} are actually associated to a different microscopic system which necessarily preserves less supersymmetry \cite{Cano:2021dyy}.\footnote{See  \cite{Chen:2024gmc} for a complementary discussion.} 

A particularly relevant case for which the $\alpha'$ corrections have not been yet analyzed is the BMPV black hole \cite{Breckenridge:1996is}, the rotating generalization of the Strominger--Vafa solution \cite{Strominger:1996sh}. The goal of this paper is precisely to fill this gap. As we further discuss below, one of the main motivations for undertaking this work is the existence of conflicting expressions for the corrected BMPV entropy in the literature. Thus, beyond the intrinsic interest of obtaining a new explicit solution, our analysis will provide an independent first-principles computation of the entropy.

\paragraph{Outline of the paper and main results.} The original BMPV solution \cite{Breckenridge:1996is} can be generalized to a supersymmetric solution of $N=2, D=5$ ungauged supergravity that carries $n_v+1$ electric charges $\{Q_I\}_{I=0, \dots, n_v}$  and one angular momentum $J_-$. This supergravity theory arises as the effective action of M-theory on a Calabi-Yau threefold CY${}_{3}$. At the two-derivative level, the couplings of the theory are determined by a fully-symmetric constant tensor $C_{IJK}$ corresponding to the intersection numbers of CY${}_3$. Additional geometrical data enter in the action beyond two derivatives. Most importantly, the coupling constant in front of the mixed Chern-Simons term $A^I\wedge {\rm Tr}(R\wedge R)$ is proportional to the second Chern class $c_{2I}$ of CY${}_{3}$.  

In this paper we shall consider a specific supergravity model which arises as a consistent truncation of heterotic string theory on $T^5$, which is dual to M-theory on $K3\times T^2$. This is the so-called STU model, which has $n_v=2$ vector multiplets and which is further determined by the choice $C_{IJK}\propto |\epsilon_{IJK}|$. Our main motivation to work with this model is that, as we already mentioned, $\alpha'$ corrections can be conveniently studied in the heterotic frame, where they are best understood. Given this, the strategy that we follow is to first embed the BMPV in heterotic theory, and then make use of the Bergshoeff-de Roo effective action \cite{Bergshoeff:1989de} in order to compute the $\alpha'$ corrections directly in ten dimensions. The corrected solution we find depends just on four functions of the radial coordinate, and it is summarized in eqs.~\eqref{eq:10Dsolution}, \eqref{eq:1form} and \eqref{eq:functions}. It should be emphasized that the solution we find interpolates between an asymptotically-flat region and the near-horizon, which is a $S^1$ fibration over AdS${}_{2}\times S^2$ as in the two-derivative solution. The corrected entropy of the solution is found in eq.~\eqref{eq:finalentropy}, which we report here for convenience:
\begin{equation}\label{eq:finalentropyintro}
 {\cal S}\,=\,2\pi \sqrt{Q_+ Q_- \left[Q_0+3-\ww^2\left(Q_0+4\right)\right]}\,.
\end{equation}
In this formula, $Q_+, Q_-, Q_0$ denote the asymptotic charges associated to momentum, winding and NS5-branes, and $\ww$ is related to the angular momentum $J_-$ by
\begin{equation}
\ww\,=\, \frac{J_-}{\sqrt{Q_+ Q_- Q_0}}\,.
\end{equation}
Before further proceeding, it should be emphasized that our final entropy formula \eqref{eq:finalentropy} holds in principle only to linear order in the large-charge expansion. At this order, one has 
\begin{equation}\label{eq:finalentropy_exp}
 {\cal S}\,=\,2\pi \sqrt{Q_+ Q_- Q_0\left(1-\ww^2\right)} \left[1+\frac{3g(\ww)}{2Q_0}\right]\,, \hspace{1cm}g(\ww)\,=\,\frac{1-\tfrac{4}{3}\ww^2}{1-\ww^2}\,.
\end{equation}
The corrected BMPV entropy has been computed in a number of references in the past literature, including \cite{Guica:2005ig, Castro:2007ci, Castro:2008ys, deWit:2009de, Gupta:2021roy, Cassani:2024tvk, Alexandrov:2026rra}. As discussed in greater detail in the last one \cite{Alexandrov:2026rra}, the previous references obtained seemingly inequivalent expressions for the function $g(\ww)$ appearing in \eqref{eq:finalentropy_exp}, even if the comparison is not always straightforward due to subtleties in the definition of conserved charges with Chern-Simons interactions, see e.g.~\cite{Cassani:2023vsa}. Nevertheless, the analysis of \cite{Alexandrov:2026rra}, which computes the microscopic 5D index of M-theory on specific classes of CY${}_3$, strongly supported the expression for the entropy presented in eq.~(7.12) of \cite{Cassani:2024tvk}. Such formula was obtained via a Legendre transform of the on-shell action of a supersymmetric non-extremal configuration corresponding to the saddle of the 5D gravitational index associated to the BMPV black hole. Remarkably, one can verify that when making the appropriate choice for the couplings of the supergravity theory considered in \cite{Cassani:2024tvk}, the formula obtained there precisely reduces to the one we find in this paper, namely \eqref{eq:finalentropyintro}. Moreover, it was found in \cite{Alexandrov:2026rra} that the match with the microscopic result improves considerably when \eqref{eq:finalentropyintro}, instead of \eqref{eq:finalentropy_exp}, is used, suggesting that \eqref{eq:finalentropyintro} might be $\alpha'$-exact. 

This provides an independent consistency check of the entropy formula originally presented in \cite{Cassani:2024tvk} working directly at the level of the heterotic effective action and only using the extremal black hole geometry. It should be stressed that the effective supergravity action considered in \cite{Cassani:2024tvk} only includes the four-derivative invariant related to the supersymmetrization of the mixed Chern-Simons term. While this term certainly arises via dimensional reduction of the heterotic effective action on a torus, it has been argued to not be enough to match the full action, see e.g.~\cite{Cai:2026qrr}. Thus, the agreement of the final entropy formulas must be explained by the fact that terms unrelated to the Chern-Simons will not contribute to the entropy of supersymmetric black holes, as explicitly verified in \cite{Cai:2026qrr} in the static case. 

\paragraph{Plan of the paper.} The paper is organized as follows. In section~\ref{sec:heterotic_theory} we review the low-energy effective action of the heterotic string at first order in $\alpha'$. In section~\ref{sec:BMPV} we embed the BMPV in the heterotic theory in ten dimensions and derive an analytic solution describing the first-order $\alpha'$ corrections. Then, in section~\ref{sec:nh+entropy} we discuss the near-horizon geometry and compute the corrected entropy. Finally, in section~\ref{sec:nonSUSYBMPV} we consider a non-supersymmetric BMPV black hole and compute the $\alpha'$ corrections as well, showing that they differ from the supersymmetric case beyond leading order in $\alpha'$. 

\paragraph{Note on conventions.} In this paper we adopt the conventions of the book \cite{Ortin:2015hya}. These coincide with those of \cite{Bergshoeff:1989de}, which will be extensively used in what follows. In particular, we shall employ the mostly-minus signature for the metric $(+-\dots -)$, and the conventions for the Riemann tensor are such that
\begin{equation}
\left[\nabla_\mu,\nabla_\nu\right]\xi^\sigma\,=\, R_{\mu\nu\rho}{}^{\sigma}\xi^\rho\, .
\end{equation}
Moreover, the Hodge dual is defined by
\begin{equation}\label{eq:Hodge}
\star(e^{a_1}\wedge \dots \wedge e^{a_p})\,=\,\frac{1}{(d-p)!}\epsilon_{b_1 \dots b_{d-p}}{}^{a_1\dots a_p}\,e^{b_1}\wedge \dots \wedge e^{b_{d-p}}\,,
\end{equation}
where $e^a$ denotes the vielbein, $d$ the number of spacetime dimensions and
\begin{equation}
\epsilon^{0\,\ldots\, d-1}\,=\,+1\,, \hspace{1cm} \Rightarrow \hspace{1cm} \epsilon_{0\,\ldots \,d-1}\,=\,\left(-1\right)^{d-1}\, . 
\end{equation}

\section{Effective action of the heterotic string at first order in $\alpha'$}\label{sec:heterotic_theory}

The supersymmetric effective action of the heterotic string at first order in $\alpha'$ can be conveniently formulated by introducing the torsionful spin connections \cite{Bergshoeff:1989de},
\begin{equation}
\omega_{(\pm)}{}^a{}_b\,=\,\omega^a{}_b \pm \frac{1}{2}H_{c}{}^a{}_b\, e^c\,,
\end{equation}
where $\omega^a{}_b$ is the Levi-Civita spin connection, $e^c$ is the zehnbein and $H$ is the Neveu-Schwarz (NS) three-form.\footnote{The first Cartan structure equation reads $\diff e^a\,=\,\omega^a{}_b\wedge e^b$.} Anomaly cancellation enforces the modified Bianchi identity \cite{Green:1984sg},
\begin{equation}\label{eq:Bianchi}
\diff H\,=\, \frac{\alpha'}{4} R_{(-)}{}^a{}_b\wedge R_{(-)}{}^b{}_a\,,
\end{equation}
where 
\begin{equation}
R_{(-)}{}^a{}_b\,=\, \diff \omega_{(-)}{}^a{}_b-\omega_{(-)}{}^a{}_c\wedge \omega_{(-)}{}^c{}_b
\end{equation}
is the curvature two-form of the torsionful spin connection $\omega_{(-)}{}^a{}_b$.

In terms of these elements, the heterotic effective action at first order in $\alpha'$ is given by \cite{Bergshoeff:1989de}\footnote{For simplicity, we are setting the 10D gauge fields to zero, which is a consistent truncation. Their effect on static black holes with three and four charges was studied in \cite{Cano:2018qev, Chimento:2018kop, Cano:2018brq}. A generalization of the BMPV black hole with non-trivial gauge fields was found in \cite{Ortin:2019sex}.}
\begin{equation}\label{eq:BdRaction}
S\,=\, \frac{g_s^2}{16\pi G_{10}} \int \diff^{10}x\,\sqrt{-g}\,e^{-2\phi}\left[R-4\left(\partial \phi\right)^2+\frac{1}{12}H^2+\frac{\alpha'}{8}R_{(-)}{}_{\mu\nu\rho\sigma}R_{(-)}{}^{\mu\nu\rho\sigma}  \right] \,,
\end{equation}
where $g_{\mu\nu}$ denotes the metric in the string frame, $\phi$ is the dilaton and $G_{10}$ is the ten-dimensional Newton constant. This is related to the asymptotic value of the string coupling $g_s$ and to $\alpha'$ by 
\begin{equation}
G_{10}\,=\,8\pi^6 g_s^2 \alpha'{}^4\,\,.
\end{equation}
The equations of motion following from the above action are \cite{Bergshoeff:1989de}\footnote{The derivation of the equations of motion simplifies after using a lemma proven in \cite{Bergshoeff:1989de}, which tells us that the variation of the action with respect to $\omega_{(-)}{}_{ab}$ is proportional to $\alpha'$ and to the two-derivative equations of motion plus terms of higher order in $\alpha'$.}
\begin{eqnarray}
\label{eq:Einstein}
 R_{\mu\nu}-2\nabla_\mu\partial_\nu \phi+\frac{1}{4}H_{\mu\rho\sigma}H_{\nu}{}^{\rho\sigma}&\,=\,&-\frac{\alpha'}{4}R_{(-)}{}_{\mu\rho\sigma\lambda}R_{(-)}{}_{\nu}{}^{\rho\sigma\lambda}\,,\\[1mm]
 \label{eq:dilaton}
 \left(\partial\phi\right)^2-\frac{1}{2}\nabla^2\phi-\frac{1}{24}H^2&\,=\,&\frac{\alpha'}{32}R_{(-)}{}_{\mu\nu\rho\sigma}R_{(-)}{}^{\mu\nu\rho\sigma}\,,\\[1mm]
 \label{eq:B}
  \diff\left(e^{-2\phi} \star H\right)&\,=\,&0\,.
\end{eqnarray}
Finally, we provide the supersymmetry transformations of the gravitino $\psi_a$ and the dilatino $\lambda$ at first order in $\alpha'$, 
\begin{eqnarray}
\label{eq:gravitino_SUSYtransf}
\delta_{\epsilon}\psi_{a}&\,=\,&\nabla^{(+)}_a\epsilon\,=\,\left(\partial_a-\frac{1}{4}\Omega_{(+)}{}_{abc}\Gamma^{bc}\right)\epsilon\,,\\[1mm]
\label{eq:dilatino_SUSYtransf}
\delta_{\epsilon}\lambda&\,=\,&\left(\Gamma^a\partial_a\phi -\frac{1}{12}H_{abc}\Gamma^{abc}\right)\epsilon\,,
\end{eqnarray}
where $\Gamma^a$ are the ten-dimensional gamma matrices.

\section{BMPV black hole at first order in $\alpha'$}
\label{sec:BMPV}

Our main goal in this section is to compute the first-order $\alpha'$ corrections to the BMPV black hole \cite{Breckenridge:1996is}. This can be seen as a solution of the STU model of $N=2, D=5$ supergravity, which can be obtained by dimensional reduction and further truncation of heterotic supergravity on a five-torus. Then, there are roughly speaking two possible strategies to compute the $\alpha'$ corrections. The first would be to reduce the effective action \eqref{eq:BdRaction} on a torus, and try to solve the resulting five-dimensional equations of motion. Alternatively, we can embed the BMPV solution in heterotic theory, and proceed by computing the $\alpha'$ corrections directly in ten dimensions. Given the complexity of the dimensionally-reduced effective action \cite{Baron:2017dvb, Elgood:2020xwu, Ortin:2020xdm, Eloy:2020dko,Jayaprakash:2024xlr}, we choose to follow the second route.

\subsection{Embedding the BMPV black hole in heterotic theory}
\label{sec:upliftBMPV}

Let us then start by uplifting the BMPV black hole to ten dimensions. To this aim, we will use the results in appendix~E.3 of \cite{RuiperezVicente:2020qfw}, where the uplift of the general class of timelike supersymmetric solutions of the STU model of $N=2, D=5$ supergravity is given. Focusing only on the subclass of these solutions which is relevant for us here, we have
\begin{eqnarray}
\diff s^2&\,=\,& \frac{1}{Z_+Z_-}\left(\diff t+\omega\right)^2-Z_0 \,\diff s^2_{\mathbb R^4}- \frac{Z_+}{Z_-}\left[\diff z- Z_+^{-1}\left(\diff t+\omega\right)\right]^2-\diff {\vec y}^2\,,\\[1mm]
\label{eq:Hsusy}
H&\,=\,&-\diff\left[Z_-^{-1}\left(\diff t+\omega\right)\wedge \diff z\right]+\star_4\diff Z_0\,,\\[1mm]
\label{eq:dilatonsusy}
e^{2\phi}&\,=\,&g_s^2 \frac{Z_0}{Z_-}\, .
\end{eqnarray}
The coordinates $z\sim z+2\pi R_z$ and ${\vec y}\sim \vec y +2\pi \sqrt{\alpha'}$ parametrize the internal space, which is a five-torus. The functions $Z_0, Z_+, Z_-$ and one-form $\omega$ are defined on the $\mathbb R^4$ \emph{base space}.\footnote{This can be in principle any hyper-Kahler space, but the one relevant for describing the BMPV solution is simply ${\mathbb R}^4$.} We introduce spherical coordinates on ${\mathbb R}^4$ such that the metric reads
\begin{equation}
\diff s^2_{\mathbb R^4}\,=\,\diff r^2 +\frac{r^2}{4} \left[(\diff \psi+\cos\theta \diff \phi)^2 + \diff \theta^2 +\sin^2\theta\diff \phi^2\right]\,,
\end{equation}
where $\theta\in [0, \pi]$ and $(\psi, \phi)\sim (\psi-2\pi, \phi+2\pi)\sim (\psi+2\pi, \phi+2\pi)$. 

At leading order in $\alpha'$, supersymmetry, equations of motion and Bianchi identities impose that the $Z_I$ functions are harmonic and that $\diff \omega$ is anti-selfdual:
\begin{equation}
\label{eq:zerothorderconditions}
\diff \star_4 \diff Z_{I}\,=\,0\,, \hspace{1cm} \diff\omega+\star_4\diff \omega\,=\,0\, ,\hspace{1cm} I=+, -,0\,.
\end{equation}
The choice corresponding to the BMPV black hole is
\begin{equation}
Z_+\,=\, 1+ \frac{q_+}{r^2}\,, \hspace{1cm} Z_-\,=\, 1+ \frac{q_-}{r^2}\,,
\hspace{1cm} Z_0\,=\, 1+ \frac{q_0}{r^2}\,,
\end{equation}
and
\begin{equation}\label{eq:omega_BMPV}
\omega\,=\,\frac{a}{2r^2}\left(\diff\psi +\cos \theta \diff \phi\right)\, .   
\end{equation}
One can verify that the one-form \eqref{eq:omega_BMPV} is indeed anti-selfdual with respect to the orientation 
\begin{equation}\label{eq:orientation}
\frac{r^3}{8}\sin\theta\,\diff r\wedge \diff \theta \wedge \diff \psi \wedge \diff \phi.
\end{equation}
As it turns out, the parameters $q_+, q_-$ and $q_0$ are related to the momentum, winding and NS5-brane charges, respectively. The precise relations can be found e.g.~in \cite{Cano:2018qev}, 
\begin{equation}\label{eq:charges}
q_+\,=\,\frac{g_s^2 \alpha'^2}{R_z^2}Q_+\,, \hspace{1cm} q_-\,=\, g_s^2\alpha' Q_-\,, \hspace{1cm} q_0\,=\,\alpha' Q_0\, .
\end{equation}
As it is well known, the BMPV black hole only carries one independent combination of the two possible angular momenta in five dimensions, $J_1, J_2$.\footnote{$J_1$ and $J_2$ generate rotations along the angular directions $\phi_1\,=\, \frac{\psi+\phi}{2}$ and $\phi_2\,=\, \frac{\phi-\psi}{2}$.} Such combination is related to the parameter $a$. Defining $J_{\pm}\,=\, \tfrac{J_1\pm J_2}{2}$, a standard calculation yields
\begin{equation}\label{eq:J_-}
 J_+\,=\,0\,, \hspace{1cm} J_-\,=\,\frac{\pi a}{4G_5} \,, 
\end{equation}
where $G_5$ is the Newton constant in five dimensions. The first condition can be understood to be a consequence of extremality, as explained in \cite{Cassani:2024tvk, Anupam:2023yns, Cassani:2024kjn, Boruch:2025qdq}. In particular, it was shown in \cite{Cassani:2024tvk} that this condition still holds after including higher-derivative corrections. We will corroborate this in the following section.

\subsection{$\alpha'$ corrections to the BMPV black hole}\label{sec:alpha_BMPV}

Based on the uplift of the supersymmetric BMPV solution in the previous section, we write down the following ansatz
\begin{eqnarray}
\diff s^2&\,=\,& \frac{1}{Z_+ Z_-}\left(\diff t+\omega\right)^2-Z_0 \,\diff s^2_{\mathbb R^4}- \frac{Z_+}{Z_-}\left[\diff z- Z_+^{-1}\left(\diff t+\omega\right)\right]^2-\diff {\vec y}^2,\\[1mm]
\label{eq:Hansatz}
H&\,=\,&-\,\diff Z_-^{-1}\wedge \left(\diff t+\omega\right)\wedge \diff z+ Z_-^{-1}\star_4\diff\omega\wedge \diff z+\star_4\diff Z_0\,,
\end{eqnarray}
where the functions $Z_{I}\,=\, Z_{I}(r)$ and the dilaton $\phi\,=\,\phi(r)$ are assumed to depend only on the radial coordinate, and
\begin{equation}
\label{eq:omega_ansatz}
\omega\,=\, W(r)\left(\diff \psi +\cos \theta \diff \phi\right)\,.   
\end{equation}
More explicitly, with respect to the orientation \eqref{eq:orientation}, we have that
\begin{equation}
\star_4\diff Z_0\,=\, \frac{r^3Z_0'}{8}\sin \theta\, \diff \theta\wedge \diff\phi\wedge \diff \psi \,,
\end{equation}
and
\begin{equation}
\star_4\diff \omega\,=\, -\frac{r W'}{2} \sin \theta \,\diff \theta\wedge \diff \phi+\frac{2W}{r}\diff r\wedge \left(\diff\psi+\cos\theta\diff\phi\right) \,.
\end{equation}
At first sight, it might seem that we have performed an unnecessary rewriting in \eqref{eq:Hansatz}. Indeed, bearing in mind the second equation in \eqref{eq:zerothorderconditions}, it is clear that \eqref{eq:Hansatz} is equivalent to \eqref{eq:Hsusy} at leading order in $\alpha'$. However, \eqref{eq:zerothorderconditions} will no longer hold at first order in $\alpha'$ if the function $W$ in \eqref{eq:omega_ansatz} receives corrections. In this case, the two ways of writing $H$ at leading order in $\alpha'$ cease to be equivalent at first order. The key point is that the sum of the first and second terms in \eqref{eq:Hansatz} is no longer closed, and this is actually what will allow us to solve the modified Bianchi identity \eqref{eq:Bianchi}, as we will see momentarily.

In what follows we explain our strategy to solve the modified Bianchi identity \eqref{eq:Bianchi} and equations of motion \eqref{eq:Einstein}, \eqref{eq:dilaton}, \eqref{eq:B}, working perturbatively in $\alpha'$. Namely, we expand around the leading order solution,
\begin{equation}\label{eq:alphaexpansion}
Z_I\,=\, 1+\frac{q_I}{r^2}+\alpha' f_I(r)\,,   \hspace{1cm} W\,=\,\frac{a}{2r^2}+\alpha' f_W(r)\,, \hspace{1cm} I\,=\, +, -, 0\,,
\end{equation}
and solve for the unknown functions $f_I$, $f_W$, and the dilaton $\phi$. As it turns out, after solving for the dilaton, the $\alpha'$-corrected equations eventually reduce to a system of inhomogeneous ODEs for the unknown functions, $f_I, f_W$. This is because the explicit $\alpha'$ terms in \eqref{eq:Bianchi}, \eqref{eq:Einstein} and \eqref{eq:dilaton} behave effectively as sources. Therefore, the functions are determined up to a solution to the homogeneous equations, which corresponds to our choice of boundary conditions, or equivalently to a shift of the constant term and the $1/r^2$ coefficient of the two-derivative solution. A natural choice of boundary conditions corresponds to fixing the asymptotic behavior of the solution such that
\begin{equation}\label{eq:bdry_cond}
Z_I\underset{r\rightarrow \infty}{\longrightarrow} 1+\frac{q_I}{r^2}+\dots\,, \hspace{1cm} W \underset{r\rightarrow \infty}{\longrightarrow} \frac{a}{2r^2}+\dots\, ,
\end{equation}
so that the relations \eqref{eq:charges} and \eqref{eq:J_-} still hold at first order in $\alpha'$. This will be our choice here. Another natural choice would be to impose mixed boundary conditions in which we fix the constant term at infinity and the coefficient of the $1/r^2$ term as $r$ approaches the position of the horizon, $r=0$. This would keep the near-horizon geometry fixed, as we further discuss in section~\ref{sec:nh+entropy}. This was the choice made e.g.~in \cite{Cano:2018qev, Faedo:2019xii, Cano:2021dyy}, where the corrections to the non-rotating three-charge solutions were obtained and further studied. Therefore, one must bear in mind this subtlety when comparing against these references. 

\paragraph{Kalb-Ramond equation.} The Kalb-Ramond equation \eqref{eq:B} imposes two conditions:
\begin{equation}\label{}
 W\left(2\phi'-\frac{Z_0'}{Z_0}+\frac{Z_-'}{Z_-} \right)\,=\,0\,,\hspace{1cm}
 2\phi'-\frac{Z_0'}{Z_0}+\frac{Z_-'}{Z_-}-\frac{Z''_-}{Z_-'}-\frac{3}{r}\,=\,0\, .
\end{equation}
In the rotating case $W\neq 0$, the system reduces to 
\begin{equation}
2\phi'-\frac{Z_0'}{Z_0}+\frac{Z_-'}{Z_-} \,=\,0\,, \hspace{1cm}
\left(r^3 Z_-'\right)'\,=\,0\, ,
\end{equation}
which is solved by
\begin{equation}
e^{2\phi}\,=\,g_s^2 \,\frac{Z_0}{Z_-}\,, 
\end{equation}
and 
\begin{equation}
 Z_-\,=\, 1+ \frac{q_-}{r^2} \,  \hspace{1cm}  \Rightarrow \hspace{1cm} f_-\,=\, 0\, .
\end{equation}
Note that we have fixed the integration constants by identifying the asymptotic value of the dilaton with the string coupling constant $g_s$, and by imposing the boundary conditions \eqref{eq:bdry_cond}.

\paragraph{Bianchi identity.} Next, we turn to the Bianchi identity. Substituting the ansatz \eqref{eq:alphaexpansion} in \eqref{eq:Bianchi} yields
\begin{equation}
\begin{aligned}
&\frac{\alpha' r^2\sin \theta}{4}\diff r\wedge \diff \theta\wedge \diff\phi\wedge \left[\left(f_0''+\frac{3}{r} f_0'-\frac{24 q_0^2 }{\left(r^2+q_0\right){}^4}\right)\,\frac{r}{2}\,\diff\psi\right. \\[1mm]
&\left. +  \left(-r \left(r^2+q_-\right) f_W''-\left(r^2+3 q_-\right) f_W'+4 r f_W+\frac{8 a q_0 r
   \left(q_0+3 q_-+4 r^2\right)}{\left(r^2+q_0\right){}^4}\right)\frac{2\diff z}{\left(r^2+q_-\right)^2}\right]\,=\,0 .
\end{aligned}
\end{equation}
The inhomogeneous term in the second line evidences that an ansatz for $H$ of the form \eqref{eq:Hsusy} would not have solved the $r\theta\phi z$ component of the Bianchi identity, since the first term in \eqref{eq:Hsusy} is closed. The above equation can be written in a simpler way by introducing new functions,
\begin{equation}
F_0\,=\,f_0-\frac{q_0^2}{r^2 \left(r^2+q_0\right){}^2}\, , \hspace{1cm}F_W\,=\,f_W-\frac{a q_0}{r^2 \left(r^2+q_0\right){}^2}\, ,
\end{equation}
namely
\begin{equation}
\frac{\alpha' r^2\sin \theta}{4}\diff r\wedge \diff \theta\wedge \diff\phi\wedge \left[\frac{\left(r^3 {F}'_0\right)'}{2r^2}\,\diff\psi- \frac{2\left(r^2+q_-\right)\left(r^3 F_W'\right)'-4r^2 \left(r^2 F_W\right)'}{r^2\left(r^2+q_-\right)^2}\diff z\right]\,=\,0 .
\end{equation}
The general solutions to these homogeneous equations are 
\begin{equation}
 F_0\,=\, c^1_0+\frac{c^2_0}{r^2}\,, \hspace{1cm} F_W\,=\,\frac{c^1_W}{r^2}+ c^2_W\left(r^2+2q_-\right)\,.   
\end{equation}
Therefore, only the trivial solutions $F_0\,=\,F_W\,=\,0$, corresponding to
\begin{equation}
f_0\,=\,\frac{q_0^2}{r^2 \left(r^2+q_0\right){}^2}\, , \hspace{1cm}f_W\,=\,\frac{a q_0}{r^2 \left(r^2+q_0\right){}^2}\, ,
\end{equation}
respect our choice of boundary conditions \eqref{eq:bdry_cond}.

\paragraph{Dilaton and Einstein equations.} As it turns out, the dilaton equation \eqref{eq:dilaton} is automatically satisfied at this point. The same occurs for all components of the Einstein equation except for the $zz$ one, which gives the following differential equation for $f_+$,
\begin{equation}
\begin{aligned}
\left(r^3 f_+'\right)'\,=\,&-\frac{16 q_- q_+ r^3 \left(q_0^2+q_0 q_-+q_-^2+3 \left(q_0+q_-\right) r^2+3
   r^4\right)}{\left(q_0+r^2\right){}^3 \left(q_-+r^2\right){}^3}\\[1mm]
&+\frac{32 a^2 r^3 \left(q_0^2+2 q_0 q_-+3 q_-^2+4 \left(q_0+2 q_-\right) r^2+6
   r^4\right)}{\left(q_0+r^2\right){}^4 \left(q_-+r^2\right){}^3}\,.
\end{aligned}
\end{equation}
The solution satisfying \eqref{eq:bdry_cond} is 
\begin{equation}
 f_+\,=\, \frac{-2}{r^2\left(r^2+q_0\right)\left(r^2+q_-\right)}\left(q_+q_--\frac{2a^2}{r^2+q_0}\right)\,.
\end{equation}

\paragraph{Summary.} We have found that the ten-dimensional configuration describing the first-order $\alpha'$ corrections to the BMPV black hole is given by 
\begin{equation}\label{eq:10Dsolution}
\begin{aligned}
\diff s^2\,=\,& \frac{1}{Z_+ Z_-}\left(\diff t+\omega\right)^2-Z_0 \,\diff s^2_{\mathbb R^4}- \frac{Z_+}{Z_-}\left[\diff z- Z_+^{-1}\left(\diff t+\omega\right)\right]^2-\diff {\vec y}^2,\\[1mm]
H\,=\,&-\,\diff Z_-^{-1}\wedge \left(\diff t+\omega\right)\wedge \diff z+ Z_-^{-1}\star_4\diff\omega\wedge \diff z+\star_4\diff Z_0\,,\\[1mm]
e^{2\phi}\,=\,& g_s^2 \frac{Z_0}{Z_-}\,,
\end{aligned}
\end{equation}
where 
\begin{equation}\label{eq:1form}
\omega\,=\,W\left(\diff \psi+\cos\theta\,\diff \phi\right)\,,
\end{equation}
and 
\begin{equation}\label{eq:functions}
\begin{aligned}
Z_0\,=\,& 1+\frac{q_0}{r^2}+ \frac{q_0^2\,\alpha'}{r^2 \left(r^2+q_0\right){}^2}+{\cal O}(\alpha'^2)\,,\\[1mm]
Z_+\,=\,& 1+\frac{q_+}{r^2}- \frac{2\alpha'}{r^2\left(r^2+q_0\right)\left(r^2+q_-\right)}\left(q_+q_--\frac{2a^2}{r^2+q_0}\right)+{\cal O}(\alpha'^2)\,,\\[1mm]
Z_-\,=\,& 1+\frac{q_-}{r^2}+{\cal O}(\alpha'^2)\,,\\[1mm]
W\,=\,& 1+\frac{a}{2r^2}+ \frac{a \,q_0\,\alpha'}{r^2 \left(r^2+q_0\right){}^2}+{\cal O}(\alpha'^2)\,.
\end{aligned}
\end{equation}
This reduces to the solution found in \cite{Cano:2018brq} when the rotation parameter $a$ is set to zero. Furthermore, in appendix~\ref{app:SUSYanalysis} we show that the corrected solution is still $\tfrac{1}{4}$-supersymmetric, as expected. 


\section{Near-horizon limit and black hole entropy}\label{sec:nh+entropy}

\subsection{Near-horizon limit}

In order to obtain the near-horizon limit of the solution we found in section~\ref{sec:alpha_BMPV}, we rescale the time and radial coordinates,
\begin{equation}
t\to \frac{t}{\lambda}\,, \hspace{1cm} r\to \sqrt{\lambda}\,r\,, 
\end{equation}
and then send $\lambda\to 0$. The resulting solution is conveniently described by introducing a new radial coordinate $\rho$ related to $r$ by
\begin{equation}
\rho\,=\, \frac{2r^2}{\sqrt{\qq_+ q_-\qq_0 -{\hat a}^2}}\,,
\end{equation}
and using the near-horizon parameters
\begin{equation}\label{eq:map_asymptotic_charges}
\qq_+\,=\,q_+\left[1-\frac{2\alpha'}{q_0}\left(1-\frac{2a^2}{q_+q_-q_0}\right)\right]\hspace{5mm} \qq_0\,=\, q_0\left(1+\frac{\alpha'}{q_0}\right)\,, \hspace{5mm}  \hat a\,=\, a\left(1+\frac{2\alpha'}{q_0}\right)\, .
\end{equation}
In terms of these quantities, the near-horizon limit of the solution reads
\begin{equation}\label{eq:metricNH}
\begin{aligned}
\diff s^2\,=\,& \frac{{\qq}_0}{4} \left(\rho^2 \diff t^2-\frac{\diff \rho^2}{\rho^2}\right)-\frac{\qq_0}{4}\left[\vv^2\left(\diff\psi+\cos\theta \diff \phi -\frac{{\hat a}\,\rho \,\diff t}{\vv\sqrt{\qq_+ q_-\qq_0 } } \right)^2+\diff\theta^2+\sin^2\theta\, \diff \phi^2\right]\\[1mm]
-&\frac{\qq_+}{\qq_-}\left[\diff z-\frac{\vv\sqrt{\qq_+ q_-\qq_0 } \,\rho\,\diff t}{2\qq_+} - \frac{\hat a}{2\qq_+}\left(\diff \psi+\cos \theta \diff \phi\right)\right]^2 \,,
\end{aligned}
\end{equation}
\begin{equation}
H\,=\,\frac{\sqrt{\qq_+ q_-\qq_0 } \, v}{2q_-}\diff t\wedge \diff\rho \wedge \diff z+\frac{\qq_0}{4}\sin\theta \,\diff\theta \wedge \diff \psi\wedge \diff \phi+\frac{\hat a}{2q_-}\sin\theta\, \diff \theta\wedge \diff \phi\wedge \diff z\,,
\end{equation}
and 
\begin{equation}\label{eq:dilatonNH}
e^{2\phi}\,=\,g_s^2 \,\frac{\qq_0}{q_-}\,.
\end{equation}
It should be noted that we are ignoring the four-torus directions $\vec y$ as they play essentially no role in our discussion. Then, in this section we will deal with the six-dimensional configuration obtained from dimensional reduction on $T^4$. From \eqref{eq:metricNH}, we see that the near-horizon geometry is the twisted product of ${\rm{AdS}}_2\times S^3\times S^1$. The squashing of the Hopf fiber is parametrized by $v$, which is given by
\begin{equation}\label{eq:squashing}
v\,=\,\sqrt{1-\frac{{\hat a}^2}{\qq_+q_-\qq_0}}\,.
\end{equation}
This is exactly the near-horizon geometry of the two-derivative solution, up to the replacement
\begin{equation}
 q_0\to \qq_0\,, \hspace{1cm}  q_+\to \qq_+\,, \hspace{1cm}   
 a\to \hat a  \,. 
\end{equation}
This implies that we can always fix our boundary conditions imposing that the near-horizon geometry is held fixed. This is not the case for us here because we decided to fix our boundary conditions at infinity ---namely, keeping the asymptotic charges fixed. An equivalent implication is that the near-horizon limit is a solution of the ``homogeneous'' or uncorrected equations of motion and Bianchi identity, meaning that it solves \eqref{eq:Bianchi}, \eqref{eq:Einstein}, \eqref{eq:dilaton} and \eqref{eq:B} with the right-hand sides set to zero. Indeed, an explicit computation shows that the near-horizon solution satisfies 
\begin{equation}\label{eq:R_-=0}
 R_{(-)}{}^{ab}\,=\,0\,,   
\end{equation}
as found earlier for static black holes with three and four charges in \cite{DominisPrester:2008ynb, Cano:2018qev, Cano:2018brq, Faedo:2019xii, Cano:2021dyy}. Another property shared with the static case is the fact that $H$ is (anti-)selfdual in six dimensions. In order to see this explicitly, we introduce the following vielbein basis,
\begin{equation}\label{eq:vielbein}
\begin{aligned}
e^0\,=\,& \frac{\sqrt{\qq_0}}{2} \rho \,\diff t\,, \hspace{2.5mm}e^1\,=\, \frac{\sqrt{\qq_0}}{2\rho}  \,\diff \rho\,, \hspace{2.5mm} e^2\,=\,\frac{\sqrt{\qq_0}}{2}\diff\theta\,,  \hspace{2.5mm} e^3\,=\, \frac{\sqrt{\qq_0} v}{2}  \left(\diff\psi+\cos\theta \diff \phi -\frac{{\hat a}\,\rho \,\diff t}{\vv\sqrt{\qq_0 \qq_+ q_-} }\right)\,, \\[1mm]
e^4\,=\,& \frac{\sqrt{\qq_0}}{2}\sin\theta\,\diff \theta\,,\hspace{5mm} e^5\,=\, \sqrt{\frac{\qq_+}{q_-}}\left[\diff z-\frac{\vv\sqrt{\qq_0 \qq_+ q_-} \,\rho\,\diff t}{2\qq_+} - \frac{\hat a}{2\qq_+}\left(\diff \psi+\cos \theta \diff \phi\right)\right]\, ,
\end{aligned}
\end{equation}
in terms of which the expression for $H$ reads
\begin{equation}\label{eq:Hflat}
H\,=\,\frac{2v}{\sqrt{\qq_0}}\left(e^0\wedge e^1\wedge e^5+e^2\wedge e^3\wedge e^4\right)+\frac{2\hat a}{\qq_0 \sqrt{\qq_+ q_-}}\left(e^0\wedge e^1\wedge e^3+e^2\wedge e^4\wedge e^5\right)\, .
\end{equation}
Therefore, in our conventions \eqref{eq:Hodge} we have that
\begin{equation}\label{eq:antiselfdualH}
H\,=\,-\star H\,.
\end{equation}

\subsection{Black hole entropy}

In this section we compute the corrected entropy of the black hole. As discussed in \cite{Tachikawa:2006sz, Elgood:2020nls}, Wald formula needs to be generalized as a consequence of the Lorentz Chern-Simons term. Following \cite{Elgood:2020nls}, we have that the generalized Wald entropy for the heterotic theory is given by
\begin{equation}\label{eq:gen_Wald_formula}
{\cal S}\,=\, \frac{g_s^2}{8 G_6}\int_{\Sigma} e^{-2\phi}\left[\star (e^a\wedge e^b)+\frac{\alpha'}{2}\star R_{(-)}{}^{ab}-\frac{\alpha'}{2}\star H\wedge \omega_{(-)}{}^{ab}\right]n_{ab}\,,
\end{equation}
where $G_6$ is the 6D Newton constant, $\Sigma$ is any spatial cross-section of the horizon\footnote{Strictly speaking, both the original formula by Wald and the one of \cite{Elgood:2020nls} are valid for non-extremal black holes, and are given in terms of integrals over the bifurcation surface. However, one can use the results of \cite{Jacobson:1993vj} to extend them to arbitrary spatial cross-sections of the horizon, rendering them applicable to extremal black holes.} and $n_{ab}$ is the binormal, normalized so that $n_{ab}\,n^{ab}=-2$. As explained in \cite{Elgood:2020nls}, the first two terms come from the variation of the Lagrangian with respect to the explicit ocurrences of the Riemann tensor, while the third one comes from the $H\wedge \star H$ term, and is not accounted for by the original Wald formula \cite{Wald:1993nt, Iyer:1994ys}.

We proceed by choosing $\Sigma$ to be a $t=$constant slice of the horizon, so that $n_{01}\,=\,1$ with respect to the vielbein basis in \eqref{eq:vielbein}. Making use of \eqref{eq:R_-=0} and of the self-duality properties for $H$ \eqref{eq:antiselfdualH}, the above formula reduces to 
\begin{equation}\label{eq:entropy}
\begin{aligned}
{\cal S}\,=\,&\frac{q_-}{4 G_6\,\qq_0}\int_{\Sigma} \left[\star (e^0\wedge e^1)+\frac{\alpha'}{2}H\wedge \omega_{(-)}{}^{01}\right]\\[2mm]
\,=\,&\frac{q_-}{4 G_6\,\qq_0}\int_{\Sigma} \left[1-\frac{\alpha'}{2}\left(H_{234}\,\omega_{(-)}{}_{501}+H_{245}\,\omega_{(-)}{}_{301}\right)\right]e^2\wedge e^3\wedge e^4\wedge e^5\\[2mm]
\,=\,&\frac{q_-}{4 G_6\,\qq_0}\int_{\Sigma} e^2\wedge e^3\wedge e^4\wedge e^5 \left(1+\frac{2\alpha'}{\qq_0}\right)\\[2mm]
\,=\,& \frac{\pi^2}{2G_5}\sqrt{\qq_+ q_- \qq_0-{\hat a}^2}\left(1+\frac{2\alpha'}{\qq_0}\right)\,.
\end{aligned}
\end{equation}
In the last line we used the relation between the Newton constants in five and six dimensions, namely $G_6\,=\,2\pi R_z \, G_5$. Additionally, in order to go from the second to the third line we used the expression for $H$ in \eqref{eq:Hflat} and that the expressions for the components of the torsionful spin connection $\omega_{(-)}{}_{abc}$ which are needed are given by
\begin{equation}\label{eq:comp_omega_-}
\omega_{(-)}{}_{501}\,=\,-\frac{2v}{\sqrt{\qq_0}}\,, \hspace{1cm}\omega_{(-)}{}_{301}\,=\,-\frac{2\hat a}{\qq_0 \sqrt{\qq_+q_-}}\, .
\end{equation}
Next, we use the map \eqref{eq:map_asymptotic_charges} in order to express the entropy in terms of the asymptotic charges,
\begin{equation}\label{eq:entropyformula1}
{\cal S}\,=\,\frac{\pi^2}{2G_5}\sqrt{q_+q_-q_0\left(1-\ww^2\right)}\left(1+\frac{3\alpha'}{2q_0}\frac{1-\tfrac{4}{3}\ww^2}{1-\ww^2}\right)\,,
\end{equation}
where, following \cite{Alexandrov:2026rra}, we have defined
\begin{equation}\label{eq:w_parameter}
\ww\,=\,\frac{a}{\sqrt{q_+ q_-q_0}}\, ,
\end{equation}
in order to facilitate the comparison with previous results in the literature Before doing so, we use \eqref{eq:charges} and \eqref{eq:J_-}, together with the fact that the 5D Newton constant is given by 
\begin{equation}
G_5\,=\,\frac{G_6}{2\pi R_z}\,=\,\frac{G_{10}}{\left(2\pi\right)^5 R_z \alpha'^2}\,=\,\frac{\pi g_s^2 \alpha'^2}{2 R_z}\,,
\end{equation}
to rewrite the entropy formula \eqref{eq:entropyformula1} in terms of $Q_+, Q_-, Q_0$ and $J_-$:
\begin{equation}\label{eq:entropyformula2}
{\cal S}\,=\,2\pi\sqrt{Q_+Q_-Q_0\left(1-\ww^2\right)}\left(1+\frac{3}{2Q_0}\frac{1-\tfrac{4}{3}\ww^2}{1-\ww^2}\right)\,, 
\end{equation}
where 
\begin{equation}
\ww\,=\,\frac{J_-}{\sqrt{Q_+ Q_- Q_0}}\,.    
\end{equation}
Finally, we note that \eqref{eq:entropyformula2} coincides at next-to-leading order with the large-charge expansion of 
\begin{equation}\label{eq:finalentropy}
{\cal S}\,=\,2\pi\sqrt{Q_+Q_-\left[Q_0+3-\ww^2\left(Q_0+4 \right)\right]}\,, 
\end{equation}
which precisely agrees with the formula presented in eq.~(7.12) of \cite{Cassani:2024tvk}, very recently reproduced via a microscopic calculation in \cite{Alexandrov:2026rra}. As shown in this reference, the agreement with the microscopic result considerably improves when the entropy is written as in \eqref{eq:finalentropy}, see Fig.~13 in \cite{Alexandrov:2026rra}. This suggests that \eqref{eq:finalentropy} might be the exact $\alpha'$-corrected entropy.

\section{Corrections in the non-supersymmetric case}
\label{sec:nonSUSYBMPV}

At leading order in $\alpha'$, the bosonic part of the heterotic effective action and equations of motion are invariant under the ${\mathbb Z}_2$ transformation,
\begin{equation}\label{eq:Z2transf}
g_{\mu\nu}\to g_{\mu\nu}\,, \hspace{1cm} H\to -H \,, \hspace{1cm} \phi \to \phi\, .
\end{equation}
Thus, if we start from a two-derivative solution and flip the sign of $H$ we will obtain another solution. Although the form of the background fields will look identical in both solutions (up to an obvious sign), they will have different physical properties. In particular, the new solution will not be supersymmetric in general, as the supersymmetric transformations are not left invariant under \eqref{eq:Z2transf}. Furthermore, the inclusion of $\alpha'$ corrections breaks the ${\mathbb Z}_2$ symmetry, and as a consequence the two solutions will differ beyond leading order in $\alpha'$. 

Given this, it is an interesting exercise to compute the first order $\alpha'$ corrections to the non-supersymmetric and extremal BMPV solution, obtained by taking the ten-dimensional solution presented in section \ref{sec:upliftBMPV} and flipping the sign of $H$. As it turns out, the same ansatz used in the previous section works in the non-supersymmetric case as well. In order to encompass both possibilities, we introduce a sign parameter $\varepsilon$, such that 
\begin{equation}
\varepsilon\,=\,+1 \hspace{2.5mm}\text{(SUSY)}\,, \hspace{1cm}\varepsilon\,=\, -1 \hspace{2.5mm}\text{(non-SUSY)}\,,
\end{equation}
as we show in appendix~\ref{app:SUSYanalysis}. The ansatz is then given by
\begin{eqnarray}
\label{eq:ansatz_metric}
\diff s^2&\,=\,& \frac{1}{Z_+ Z_-}\left(\diff t+\omega\right)^2-Z_0 \,\diff s^2_{\mathbb R^4}- \frac{Z_+}{Z_-}\left[\diff z- Z_+^{-1}\left(\diff t+\omega\right)\right]^2-\diff {\vec y}^2,\\[1mm]
\label{eq:ansatz_H}
H&\,=\,&\varepsilon\left[-\,\diff Z_-^{-1}\wedge \left(\diff t+\omega\right)\wedge \diff z+ Z_-^{-1}\star_4\diff\omega\wedge \diff z+\star_4\diff Z_0\right]\,,\\[1mm]
\label{eq:ansatz_phi}
e^{2\phi}&\,=\,&g_s^2 \frac{Z_0}{Z_-}\,,
\end{eqnarray}
where, as before,
\begin{equation}
\omega\,=\, W\left(\diff \psi +\cos \theta \diff \phi\right)\,, 
\end{equation}
and all the functions of the ansatz are assumed to depend only on the radial coordinate. Following the same steps as in the supersymmetric case, we obtain the $\alpha'$ corrections to the harmonic functions are given by
\begin{equation}
\begin{aligned}
Z_0\,=\,& 1+\frac{q_0}{r^2}+ \frac{q_0^2\,\alpha'}{r^2 \left(r^2+q_0\right){}^2}+{\cal O}(\alpha'^2)\,,\\[1mm]
Z_+\,=\,& 1+\frac{q_+}{r^2}- \frac{\left(1+\varepsilon\right)\alpha'}{r^2\left(r^2+q_0\right)\left(r^2+q_-\right)}\left(q_+q_--\frac{2a^2}{r^2+q_0}\right)+{\cal O}(\alpha'^2)\,,\\[1mm]
Z_-\,=\,& 1+\frac{q_-}{r^2}+{\cal O}(\alpha'^2)\,,\\[1mm]
W\,=\,& 1+\frac{a}{2r^2}+ \frac{\left(1+\varepsilon\right) a \,q_0\,\alpha'}{2r^2 \left(r^2+q_0\right){}^2}+{\cal O}(\alpha'^2)\,.
\end{aligned}
\end{equation}
As found in the static case \cite{Cano:2021nzo, Ortin:2021win}, we find that the $\alpha'$ corrections actually simplify in the non-supersymmetric case, as only the function $Z_0$ receives corrections.

\paragraph{Non-supersymmetric black hole entropy.} The metric and dilaton in the near-horizon limit take the same form as in the supersymmetric solution, \eqref{eq:metricNH} and \eqref{eq:dilatonNH}, but bearing in mind that the relation between the near-horizon and asymptotic charges \eqref{eq:map_asymptotic_charges} is modified by $\varepsilon$. Indeed, we find that in the non-supersymmetric case
\begin{equation}\label{eq:nonSUSY_map_asymptotic_charges}
\qq_+\,=\,q_+\,, \hspace{1cm} \qq_0\,=\,q_0\left(1+\frac{\alpha'}{q_0}\right)\,, \hspace{1cm} \hat a\,=\, a\, ,\hspace{1cm} \varepsilon\,=\,-1 \hspace{2.5mm}\text{(non-SUSY)}\,.
\end{equation}
Additionally, the three-form picks an overall sign with respect to the expression in the supersymmetric case \eqref{eq:Hflat}. Crucially, this sign implies that the two components of the torsionful spin connection entering in the formula for the entropy \eqref{eq:entropy} vanish. As a consequence, the entropy in the non-supersymmetric case is just given by the area term
\begin{equation}\label{eq:nonSUSYentropy}
{\cal S}_{\text{non-susy}} \,=\,\frac{\pi^2}{2G_5}\sqrt{q_+q_-\qq_0-a^2}\,=\,\frac{\pi^2}{2G_5}\sqrt{q_+q_-q_0\left(1-\ww^2\right)}\left(1+\frac{\alpha'}{2q_0}\right)\,,
\end{equation}
where we recall that the parameter $\ww$ was defined in \eqref{eq:w_parameter}. This result reproduces the one obtained in \cite{Cano:2021nzo} in the static case, $\ww\,=\,0$. Finally, it should be emphasized that we have obtained the corrections in the non-supersymmetric case just as an academic exercise. Starting with \cite{Iliesiu:2020qvm}, it has been understood in recent years that non-supersymmetric near-extremal black holes receive large quantum corrections. These modify the density of states, driving it to zero in the extremal limit. Consequently, the above formula for the entropy cannot be really trusted.

\section*{Acknowledgments}
I would like to thank Davide Cassani, Tom\'as Ort\'in and Enrico Turetta for interesting discussions and comments on the draft. I am supported by the University of Padua and the Fondazione Cariparo under the STARS@UNIPD 2025 programme (GRASBH -- The gravitational path integral, supersymmetric black holes and higher-derivative corrections).

\appendix

\section{Supersymmetry analysis}
\label{app:SUSYanalysis}
Let us consider the field configuration given in \eqref{eq:ansatz_metric}, \eqref{eq:ansatz_H}, \eqref{eq:ansatz_phi} and make the following choice of zehnbein basis $e^a$, 
\begin{equation}
\begin{aligned}
e^0\,=\,&\frac{1}{\sqrt{Z_+ Z_-}}\left(\diff t+\omega\right)\,, \hspace{2.5mm} e^1\,=\,\sqrt{Z_0}\,\diff r\,,  \hspace{2.5mm} e^2\,=\,\frac{r\sqrt{Z_0}}{2}\,\diff \theta\,, \hspace{2.5mm} e^3\,=\,\frac{r\sqrt{Z_0}}{2}\,\left(\diff \psi+\cos\theta\diff \phi\right)\,,\\[1mm]
e^4\,=\,&\frac{r\sqrt{Z_0}}{2}\sin \theta\,\diff \phi\,, \hspace{2.5mm} e^5\,=\,\sqrt{\frac{Z_+}{Z_-}}\left[\diff z- Z_+^{-1}\left(\diff t+\omega\right)\right]\,,  \hspace{2.5mm} e^{i+6}\,=\,\diff y^i\,,
\end{aligned}
\end{equation}
where $i=1, \dots, 4$ are the four-torus directions. In this basis, the dilatino Killing spinor equation (KSE), 
\begin{equation}
\delta_{\epsilon}\lambda\,=\,\left(\Gamma^a\partial_a\phi -\frac{1}{12}H_{abc}\Gamma^{abc}\right)\epsilon\,=\,0\, , 
\end{equation}
is satisfied for arbitrary choices of the functions $Z_+, Z_-, Z_0, W$ if the following conditions are met:
\begin{equation}
\left(1-\varepsilon\Gamma^{1234}\right)\epsilon\,=\,0\,,\hspace{1cm} \left(1+\varepsilon\Gamma^{05}\right)\epsilon\,=\,0\,,\hspace{1cm} \left(1+\Gamma^{05}\right)\epsilon\,=\,0\,.
\end{equation}
We note that the last one is needed only if $W\neq0$. In this case, the dilatino KSE already informs us that only the configuration with $\varepsilon\,=\,+1$ can be supersymmetric, since we cannot impose the two conditions $\left(1\pm\Gamma^{05}\right)\epsilon\,=\,0$ simultaneously, unless $\epsilon=0$. 

Thus, from now on we assume $\varepsilon\,=\,1$ and that the Killing spinor satisfies
\begin{equation}\label{eq:KS_cond}
\left(1-\Gamma^{1234}\right)\epsilon\,=\,0\,,\hspace{1cm} \left(1+\Gamma^{05}\right)\epsilon\,=\,0\,.
\end{equation}
Under these assumptions, the gravitino KSE,
\begin{equation}
\delta_\epsilon \psi_a\,=\,\left(\partial_a-\frac{1}{4}\Omega_{(+)}{}_{abc}\Gamma^{bc}\right)\epsilon\,=\,0\,, 
\end{equation}
boils down to the following first-order differential equations,
\begin{eqnarray}
\partial_t\epsilon&\,=\,&\partial_\psi\epsilon\,=\,\partial_z\epsilon\,=\,\partial_{y^i}\epsilon\,=\,0\,,\\[1mm]
\label{eq:radialeq}
\partial_r\epsilon&\,=\,&-\frac{1}{4}\left(\frac{Z_+'}{Z_+}+\frac{Z_-'}{Z_-}\right)\epsilon\,,\\[1mm]
\label{eq:thetaeq}
\partial_\theta\epsilon&\,=\,&-\frac{\Gamma^{12}}{2}\epsilon\,,\\[1mm]
\label{eq:phieq}
\partial_\phi\epsilon&\,=\,&-\frac{1}{2}\left(\sin \theta \,\Gamma^{14}+\cos\theta \,\Gamma^{13}\right)\epsilon\,=\,-\frac{1}{2}\left(-\sin \theta \,\Gamma^{12}+\cos\theta \,\right)\Gamma^{13}\epsilon\,.
\end{eqnarray}
To solve this system of equations, we make an ansatz of the form
\begin{equation}\label{eq:ansatzKS}
\epsilon\,=\, \epsilon_r(r)M(\theta)N(\phi)\epsilon_0\,,
\end{equation}
where $\epsilon_r(r)$ is a function, $M(\theta)$ and $N(\phi)$ are matrices ---which are assumed to commute with the projectors in \eqref{eq:KS_cond}--- and $\epsilon_0$ is a constant spinor satisfying \eqref{eq:KS_cond}, namely
\begin{equation}\label{eq:KS_cond_0}
\left(1-\Gamma^{1234}\right)\epsilon_0\,=\,0\,,\hspace{1cm} \left(1+\Gamma^{05}\right)\epsilon_0\,=\,0\,.
\end{equation}
Then we can straightforwardly integrate the radial equation \eqref{eq:radialeq}, obtaining that
\begin{equation}
 \epsilon_r\,=\, \left(Z_+Z_-\right)^{-\tfrac{1}{4}}  \,.  
\end{equation}
Next we plug \eqref{eq:ansatzKS} in \eqref{eq:thetaeq}, which yields the following matrix differential equation,
\begin{equation}
\left(\frac{\diff M(\theta)}{\diff \theta}+\frac{1}{2}\Gamma^{12}M(\theta) \right)\,N(\phi)\,\epsilon_0\,=\,0\,,
\end{equation}
which is solved by 
\begin{equation}
M(\theta)\,=\,e^{-\frac{1}{2}\Gamma^{12}\theta}\,=\,\cos\left(\tfrac{\theta}{2}\right)-\sin\left(\tfrac{\theta}{2}\right)\Gamma^{12}\,.  
\end{equation}
This satisfies the property,
\begin{equation}
\left(\cos \theta-\sin \theta \Gamma^{12}\right)\Gamma^{13}M(\theta)\,=\,M(\theta)\Gamma^{13}\,, 
\end{equation}
which can be used in \eqref{eq:phieq} to find that $N(\phi)$ satisfies the differential equation, 
\begin{equation}
 \frac{\diff N(\phi)}{\diff \phi} + \frac{1}{2}\Gamma^{13}N(\phi)\,=\,0\,, 
\end{equation}
so that 
\begin{equation}
N(\phi)\,=\,e^{-\frac{1}{2}\Gamma^{13}\phi}\,=\,\cos\left(\tfrac{\phi}{2}\right)-\sin\left(\tfrac{\phi}{2}\right)\Gamma^{13}\,.
\end{equation}
Thus, the Killing spinor is given by 
\begin{equation}
\epsilon\,=\,\frac{1}{\left(Z_+Z_-\right)^{1/4}}\left[\cos\left(\tfrac{\theta}{2}\right)-\sin\left(\tfrac{\theta}{2}\right)\Gamma^{12}\right]\left[\cos\left(\tfrac{\phi}{2}\right)-\sin\left(\tfrac{\phi}{2}\right)\Gamma^{13}\right]\epsilon_0\,,
\end{equation}
where we recall that $\epsilon_0$ is a constant spinor satisfying the two conditions in \eqref{eq:KS_cond_0}, which imply that field configuration preserves one-quarter of the total number of supersymmetries.
\bibliography{refs.bib}
\bibliographystyle{JHEP}

\end{document}